\documentclass{ws-mpla}

\begin{document}

\markboth{Kalam {\em et al}}
{A relativistic model for strange quark stars}

%%%%%%%%%%%%%%%%%%%%% Publisher's Area please ignore %%%%%%%%%%%%%%
\catchline{}{}{}{}{}
%%%%%%%%%%%%%%%%%%%%%%%%%%%%%%%%%%%%%%%%%%%%%%%%%%%%%%%%%%%%%%%%%%%

\title{A RELATIVISTIC MODEL FOR STRANGE QUARK STARS}

\author{\footnotesize MEHEDI KALAM\footnote{E-mail: kalam@iucaa.ernet.in}}
\address{Department of Physics, Aliah University, Kolkata 700091, West Bengal, India.}

\author{\footnotesize ANISUL AIN USMANI\footnote{E-mail: anisul@iucaa.ernet.in}}
\address{Department of Physics, Aligarh Muslim University, Aligarh 202 002, Uttar Pradesh, India.}

\author{\footnotesize FAROOK RAHAMAN\footnote{E-mail: rahaman@iucaa.ernet.in}}
\address{Department of Mathematics, Jadavpur University, Kolkata 700 032, West Bengal, India.}

\author{\footnotesize Sk. MONOWAR HOSSEIN\footnote{E-mail: sami\_milu@yahoo.co.uk}}
\address{Department of Mathematics, Aliah University, Kolkata 700091, West Bengal, India.}

\author{\footnotesize INDRANI KARAR\footnote{E-mail: indrani.karar08@gmail.com}}
\address{Department of Mathematics, Saroj Mohan Institute of Technology, Guptipara, West
Bengal, India.}

\author{\footnotesize RANJAN SHARMA\footnote{E-mail: rsharma@iucaa.ernet.in}}
\address{Department of Physics, P. D. Women's College, Jalpaiguri 735101, India.}

\maketitle

\pub{Received (Day Month Year)}{Revised (Day Month Year)}

\begin{abstract}
We propose a spherically symmetric and anisotropic model for strange quark stars within the framework of MIT Bag model. Though the model is found to comply with all the physical requirements of a realistic star satisfying a strange matter equation of state (EOS), the estimated values the Bag constant for different strange star candidates like Her X-1, SAX J 1808.4-3658 and 4U 1820-30, clearly indicate that the Bag constant need not necessarily lie within the range of $60-80$~MeV~fm$^{-3}$ as claimed in the literature\cite{Farhi1984,Alcock1986}.
\keywords{General relativity; Exact solutions; Strange stars.}
\end{abstract}

\ccode{PACS Nos.: 04.40.Nr, 04.20.Jb, 04.20.Dw}

\section{Introduction}
In 1964, Gell-Mann\cite{GellMann1964} and Zweig\cite{George1964} had independently suggested that hadrons are composed of even more fundamental particles called quarks, a proposition which got experimental support later on. In the theoretical front, the quark matter hypothesis put forward by Witten\cite{Witten}, had prompted many investigators to investigate an entirely new class of compact astrophysical objects composed of strange quark matter called strange stars (see Weber\cite{Weber2005} for a recent review). As quarks are not seen as free particles, the quark confinement mechanism have been dealt with great details in QCD. In the MIT bag model\cite{Chodos1974} for strange stars, the quark confinement has been assumed to be caused by a universal pressure $B$, called the bag constant. Farhi and Jafee\cite{Farhi1984} and Alcock {\em et al}\cite{Alcock1986} had shown that for a stable strange quark matter the value of the bag constant should be $B \sim 55 - 75~$MeV~fm$^{-3}$.

At the backdrop of such theoretical developments, analytic model building of strange quark stars naturally plays a crucial role to understand the gravitational behaviour of strange stars. In this vein, from the perspective of classical GR, we propose here an analytic solution capable of describing realistic strange stars. In an earlier work, some of us\cite{Farook2012} had proposed a relativistic model for strange stars where the geometric model of Krori and Barua\cite{Krori1975} was used to describe the interior space-time of strange stars. In this paper, we adopt the Finch and Skea\cite{Finch1989} model to develop the interior space-time of a strange star. We assume a spherically symmetric star composed of strange quark matter whose equation of state (EOS) is  governed by the MIT bag model. We also assume anisotropy in the matter composition implying that the radial pressure ($p_r$) is not equal to the tangential pressure ($p_t$). Since the density of a strange star may exceed the nuclear matter density it is expected that pressure at the interior of such would be anisotropic, in general\cite{Bowers1974,Soko1980}. Anisotropy in a  highly dense compact stellar object may occur for various reasons (e.g, existence of a solid core, phase transition, presence of electromagnetic field etc.) whose contributions in the stellar modelling can be incorporated by assuming  the star to be anisotropic in general\cite{Ivanov2010}. In our construction, the assumption of anisotropy provides an extra degree of freedom to deal with the system.

The strange star model, in this work, has been developed by considering the MIT bag model EOS and a particular ansatz for the metric function $g_{rr}$ proposed by Finch and Skea\cite{Finch1989}. The solution found here has been shown to comply with all the physical requirements of a realistic star. When applied to some proposed strange star candidates, it has been found that the corresponding values of the bag constant lie on the higher side as compared to its acceptable range put forward by Farhi and Jafee\cite{Farhi1984} and Alcock {\em et al}\cite{Alcock1986}. In fact, our results are consistent with the experimental results from CERN-SPS and RHIC, indicating a wider range of values of the bag constant\cite{Burgio}.

\section{Interior solution:}
We assume that the interior space-time of a strange star is described by a spherically symmetric metric of the form
\begin{equation}
ds^2 = -e^{\nu(r)}dt^2 +  \left(1+\frac{r^2}{R^2}\right)dr^2 +r^2
(d\theta^2 +sin^2\theta d\phi^2), \label{eq1}
\end{equation}
where, the metric function $\nu(r)$ is yet to be determined. Note that the ansatz for the metric function $g_{rr}$ in (\ref{eq1}) was proposed by Finch and Skea\cite{Finch1989} to develop a viable model for a relativistic compact star. The $t=constant$ hyper-surface of the metric is paraboloidal in nature which exhibits a departure from spherical geometry and the constant $R$ is a curvature parameter which governs the geometry of the back ground space-time.

We assume that the energy-momentum tensor for the strange matter filling the interior of the star has the standard form $T_{ij} = diag(\rho,-p_r,-p_t,-p_t)$, where, $\rho$ is the energy-density; $p_r$ and $p_t$ are the
radial and transverse pressures, respectively. Einstein's field equations for the line-element (\ref{eq1}), accordingly, are  obtained as (we set $G = c=1$)
\begin{eqnarray}
8\pi\rho &=& \frac{1}{R^2}\left(3+\frac{r^2}{R^2}\right)\left(1+\frac{r^2}{R^2}\right)^{-2}, \label{eq2}\\
8\pi p_r  &=& \left(1+\frac{r^2}{R^2}\right)^{-1}\left[\frac{\nu^\prime}{r}+\frac{1}{r^2}\right] -
\frac{1}{r^2}, \label{eq3}\\
8\pi p_t &=& \left(1+\frac{r^2}{R^2}\right)^{-1}
\left[\frac{\nu^{\prime\prime}}{2}+\frac{\nu^\prime}{2r}+\frac{{\nu^\prime}^2}{4}\right] -\frac{1}{R^2}\left(1+\frac{r^2}{R^2}\right)^{-2}\left[1+\frac{\nu^\prime r}{2}\right].\label{eq4}
\end{eqnarray}
Eqs.~(\ref{eq2})-(\ref{eq4}) constitute a system of four unknowns ($\rho$, $p_r$, $p_t$, $\nu$). By suitably choosing any one of these unknown parameters, the system may be solved. Our objective here is to develop a model for strange stars and, therefore, we assume the simplest form of the strange matter EOS having the form
\begin{equation}
p_r = \frac{1}{3}(\rho-4B),\label{eq5}
\end{equation}
where, $B$ is the bag constant. Substituting Eq.~(\ref{eq5}) in Eq.~(\ref{eq3}) and integrating,  we determine the unknown metric function $\nu$ in the form
\begin{equation}
\nu = \frac{1}{3}\ln(R^2+r^2)-\frac{8\pi
B}{3}\frac{r^2}{R^2}\left(2R^2+r^2-\frac{1}{4\pi B}\right) +
\nu_0.\label{eq6}
\end{equation}
where $\nu_0 $ is an integration constant.

 Using Eq.~(\ref{eq6})
in Eqs.~(\ref{eq2})-(\ref{eq4}), the density and the two pressures
are then obtained as
\begin{eqnarray}
8\pi\rho &=& \frac{1}{ R^2}\frac{\left(3+\frac{r^2}{R^2}\right)}{\left(1+\frac{r^2}{R^2}\right)^2},\label{eq7}\\
8\pi p_{r} &=& \frac{1}{3 R^2}\frac{\left(3+\frac{r^2}{R^2}\right)}{\left(1+\frac{r^2}{R^2}\right)^2}-\frac{32\pi}{3}
B,\label{eq8}\\
8\pi p_{t} &=& \left(1+\frac{r^2}{R^2}\right)^{-1} \left[\frac{256\pi^2 B^2R^2}{9}\left(\frac{r}{R}\right)^6 +\left(\frac{512\pi^2
B^2R^2}{9}-\frac{64\pi B}{9}-\frac{32\pi~B}{9(1+\frac{r^2}{R^2})}\right)\left(\frac{r}{R}\right)^4 \right.\nonumber\\
&& +\left(\frac{256 \pi^2 B^2R^2}{9}-\frac{64 \pi B}{9}+\frac{4}{9R^2} -\frac{2}{9R^2
(1+\frac{r^2}{R^2})^2}+\frac{1}{R^2(1+\frac{r^2}{R^2})}-\frac{32\pi B}{9(1+\frac{r^2}{R^2})}+\frac{4}{3r^2}\right)\left(\frac{r}{R}\right)^2\nonumber \\
&& \left. +\frac{1}{3R^2(1+\frac{r^2}{R^2})^2} +\frac{1}{3R^2(1+\frac{r^2}{R^2})}-\frac{32\pi B}{3}\right]
-\left(1+\frac{r^2}{R^2}\right)^{-2}\left[\frac{2}{3R^2}\left(\frac{3}{2}+\frac{r^2}{R^2}\right)\right.\nonumber\\
&& \left.+\frac{1}{3R^2}\frac{(\frac{r}{R})^2}{\left(1+\frac{r^2}{R^2}\right)}-\frac{16\pi
B}{3}\left(\frac{r}{R}\right)^2\left(1+\frac{r^2}{R^2}\right)\right].\label{eq9}
\end{eqnarray}

\section{Physical behaviour of the model}
We note that the physical behaviour of the model depends on the constants $R$ and $B$. We need to put appropriate bounds on these parameters so that the model can describe a realistic strange star. To this end, based on various physical requirements, let us now analyze the behaviour of the physical parameters.

\begin{figure}[ptb]
\begin{center}
\vspace{0.3cm}\includegraphics[width=0.5\textwidth]{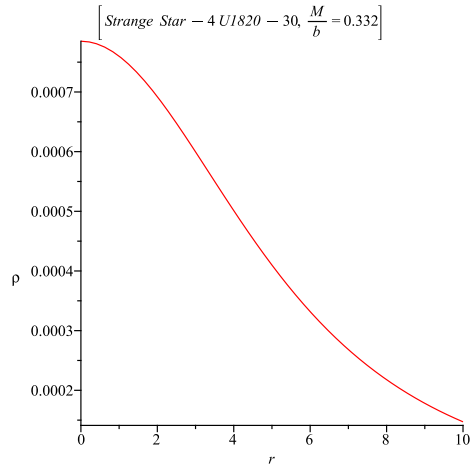}
\end{center}
\caption{Variation of the energy-density ($\rho$) at the interior
of the star.}
\label{fig:1}
\end{figure}

\begin{figure}[ptb]
\begin{center}
\vspace{0.3cm}\includegraphics[width=0.5\textwidth]{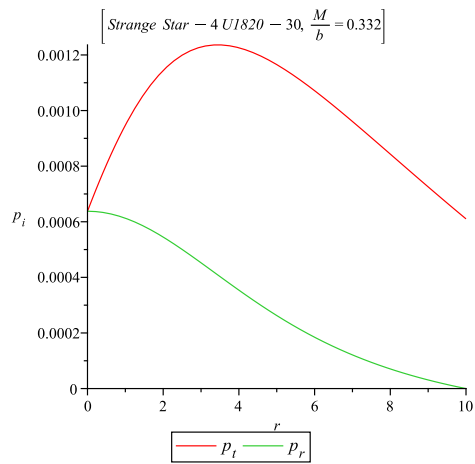}
\end{center}
\caption{Variation of the radial($p_r$) and transverse pressure($p_t$) at the interior of the star.}
\label{fig:2}
\end{figure}

Let us assume that $b$ be the radius of the star. Then, from Eq.~(\ref{eq8}), the central and surface densities are respectively obtained as
\begin{eqnarray}
\rho_0 &=& \frac{3}{8\pi R^2},\label{eq10}\\
\rho_b &=& \frac{1}{8\pi
R^2}\left(3+\frac{b^2}{R^2}\right)\left(1+\frac{b^2}{R^2}\right)^{-2}.\label{eq11}
\end{eqnarray}
We also have
\begin{eqnarray}
\frac{d\rho}{dr} &=& - \frac{r(5+\frac{r^2}{R^2})}{4\pi R^4(1+\frac{r^2}{R^2})^3}< 0,\label{eq12}\\
\frac{d\rho}{dr} (r=0) &=& 0,\label{eq13}\\
\frac{d^2 \rho}{dr^2}(r=0) &=& -\frac{5}{4\pi R^4} < 0.\label{eq14}
\end{eqnarray}
Obviously, the density is maximum at the centre and it decreases radially outward. Similarly, from Eq.~(\ref{eq8}), we have
\begin{eqnarray}
\frac{dp_r}{dr} &=& - \frac{r(5+\frac{r^2}{R^2})}{12\pi R^4(1+\frac{r^2}{R^2})^3} < 0,\label{eq15}\\
\frac{dp_r}{dr}(r=0) &=& 0,\label{eq16}\\
\frac{d^2 p_r}{dr^2} &=& -\frac{5}{12\pi R^4} < 0,\label{eq17}
\end{eqnarray}
which show that the radial pressure also decreases from the centre towards the boundary. Thus, the energy density and the radial
pressure are well behaved in the interior of the stellar configuration. Variations of the energy-density and two pressures
have been shown in Fig.~(\ref{fig:1}) and (\ref{fig:2}), respectively.

\begin{figure}[ptb]
\begin{center}
\vspace{0.3cm}\includegraphics[width=0.5\textwidth]{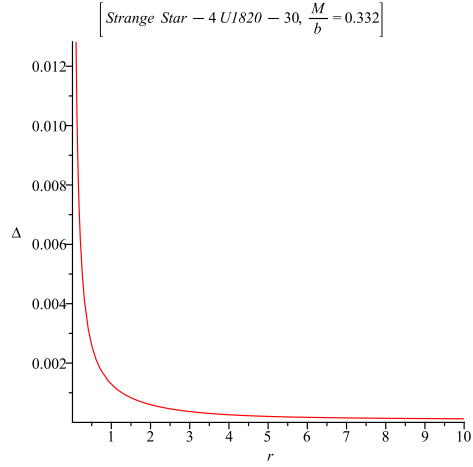}
\end{center}
\caption{Variation of the anisotropic stress $\Delta (r)  =
\frac{2}{r}\left(p_t-p_r\right)$ at the stellar interior.}
\label{fig:3}
\end{figure}

The anisotropic parameter $\Delta (r)  = \frac{2}{r}\left(p_t-p_r\right)$ is obtained as
\begin{eqnarray}
\Delta  = \frac{1}{4\pi r}\left(1+\frac{r^2}{R^2}\right)^{-1}\left[\frac{256
\pi^2B^2R^2}{9}\left(\frac{r}{R}\right)^6 +\left(\frac{512
\pi^2B^2R^2}{9}-\frac{64\pi
B}{9}-\frac{32\pi~B}{9(1+\frac{r^2}{R^2})}\right)\left(\frac{r}{R}\right)^4\right.\nonumber
\\
+\left(\frac{256 \pi^2B^2R^2}{9} - \frac{64 \pi B}{9}+\frac{4}{9R^2}-\frac{2}{9R^2(1+\frac{r^2}{R^2})^2}
 +\frac{1}{R^2(1+\frac{r^2}{R^2})}-\frac{32\pi B}{9(1+\frac{r^2}{R^2})}+\frac{4}{3r^2}\right)\left(\frac{r}{R}\right)^2\nonumber
\\
\left.+\frac{1}{3R^2(1+\frac{r^2}{R^2})^2}+\frac{1}{3R^2(1+\frac{r^2}{R^2})}-\frac{32\pi
 B}{3}\right]\nonumber
\\
 -\frac{1}{4\pi r}\left(1+\frac{r^2}{R^2}\right)^{-2}\left[\frac{2}{3R^2}\left(\frac{3}{2}+\frac{r^2}{R^2}\right)
+\frac{1}{3R^2}\frac{\left(\frac{r}{R}\right)^2}{(1+\frac{r^2}{R^2})}
-\frac{16\pi
B}{3}\left(\frac{r}{R}\right)^2(1+\frac{r^2}{R^2})\right]\nonumber\\
-\frac{1}{12 \pi r
R^2}\frac{(3+\frac{r^2}{R^2})}{(1+\frac{r^2}{R^2})^2}+\frac{8B}{3r}.\label{eq18}
\end{eqnarray}
Figure (\ref{fig:3}) shows the nature of the anisotropic stress at the stellar interior for a particular case. The condition that the anisotropic parameter $\Delta$ should vanish at the centre ($r=0$) yields
\begin{equation}
\frac{1}{8\pi R^2}- \frac{4}{3}B = 0,\label{eq19}
\end{equation}
which can be used to calculate the Bag constant $B$.

\subsection{Matching conditions}
At the boundary of the star $r=b$, the interior metric should be matched to the Schwarzschild exterior metric. Continuity of the
metric functions across the boundary surface yields
\begin{eqnarray}
\left(1+\frac{b^2}{R^2}\right)^{-1} &=& 1 - \frac{2M}{b},\label{eq20}\\
\nu(r=b) &=& \ln\left(1-\frac{2M}{b}\right) = \frac{1}{3}\ln(R^2+b^2)\nonumber \\
&&-\frac{8\pi B}{3}\frac{b^2}{R^2}\left(2R^2+b^2-\frac{1}{4\pi
B}\right).\label{eq21}
\end{eqnarray}
From Eqs.~(\ref{eq20}), we obtain the compactness of the star as
\begin{equation}
\frac{M}{b} =
\frac{b^2}{2R^2}\left(1+\frac{b^2}{R^2}\right)^{-1}.\label{eq22}
\end{equation}

The condition that the radial pressure must vanish at the boundary ($p_r( r=b) = 0$) determines the  bag constant in the form
\begin{equation}
B = \frac{1}{32 \pi R^2}
\frac{\left(3+\frac{b^2}{R^2}\right)}{\left(1+\frac{b^2}{R^2}\right)^2}.\label{eq23}
\end{equation}

\subsection{TOV equation}
\begin{figure}[ptb]
\begin{center}
\vspace{0.3cm}\includegraphics[width=0.5\textwidth]{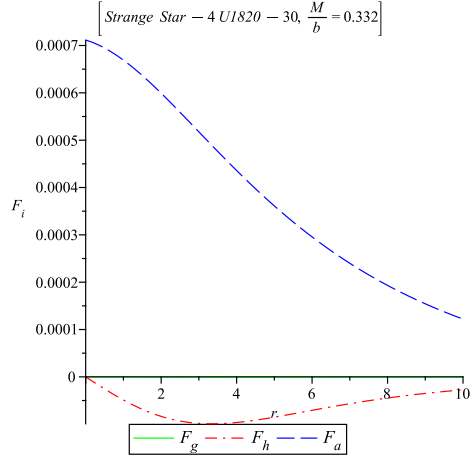}
\end{center}
\caption{Three different forces acting on the fluid elements of a given configuration.}
\label{fig:4}
\end{figure}

For an anisotropic fluid distribution, the generalized Tolman-Oppenheimer-Volkoff (TOV) equation gets the form
\begin{equation}
\frac{dp_r }{dr} +\frac{1}{2} \nu^\prime\left(\rho +p_r\right) +
\frac{2}{r}\left(p_r - p_t\right) = 0.\label{eq24}
\end{equation}
We rewrite the above equation in the form
\begin{equation}
-\frac{M_G\left(\rho+p_r\right)}{r^2}e^{\frac{\lambda-\nu}{2}}-\frac{dp_r
}{dr}
 +\frac{2}{r}\left(p_t-p_r\right) = 0, \label{eq25}
\end{equation}
where, $M_G(r)$ is the effective gravitational mass inside a
sphere of radius $r$ and is given by
\begin{equation}
M_G(r) = \frac{1}{2}r^2e^{\frac{\nu-\lambda}{2}}\nu^{\prime}.\label{eq26}
\end{equation}
The modified TOV equation describes the equilibrium condition for
the strange star subject to gravitational ($F_g$), hydrostatic ($F_h$) and anisotropic ($F_a$) stresses within the stellar interior so that
\begin{equation}
F_g+ F_h + F_a = 0,\label{eq27}
\end{equation}
where the stress components are given by
\begin{eqnarray}
F_g &=& -B r\left(\rho+p_r\right),\label{eq28}\\
F_h &=& -\frac{dp_r}{dr} ,\label{eq29}\\
F_a &=& \frac{2}{r}\left(p_t -p_r\right).\label{eq30}
\end{eqnarray}
In Fig.~\ref{fig:4}, we have shown variations of $F_g$, $F_h$ and $F_a$ for a particular stellar configuration. Obviously, it is possible to construct a static equilibrium configuration in the presence of anisotropic, gravitational and hydrostatic  stresses.

\subsection{Energy conditions}
Imposition of the energy conditions, namely, the null (NEC), weak (WEC), strong (SEC) and dominant (DEC) energy conditions, put certain bounds on the model parameters. In our model, applying these energy conditions at the
centre ($r=0$), we get the following bounds:\\
(i) NEC: $p_0+\rho_0\geq0$ $\Rightarrow$ $B\leq \frac{3}{8\pi R^2}$, i.e. $B\leq \rho_0$,\\
(ii) WEC: $p_0+\rho_0\geq0$ $\Rightarrow$  $B\leq \frac{3}{8\pi R^2}$, i.e. $B\leq \rho_0$, $\rho_0\geq0$,\\
(iii) SEC: $p_0+\rho_0\geq0$  $\Rightarrow$  $B\leq \rho_0$, $3p_0+\rho_0\geq0$ $\Rightarrow$ $B\leq \frac{\rho_0}{2}$,\\
(iv) DEC: $\rho_0 > |p_0|$ $\Rightarrow$ $B\leq \frac{\rho_0}{2}$.\\
Values of the model parameters for different stellar configurations are in agreement with these bounds as shown in Table $2$.

\subsection{Mass-Radius relation}
\begin{figure}[ptb]
\begin{center}
\vspace{0.3cm}\includegraphics[width=0.5\textwidth]{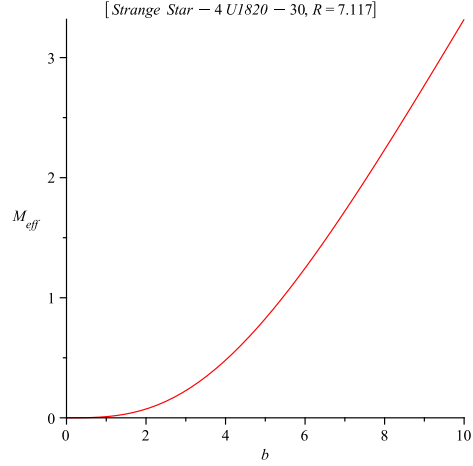}
\end{center}
\caption{Variation of $M_{eff} $ against radial parameter $r$.}
\label{fig:5}
\end{figure}

\begin{figure}[ptb]
\begin{center}
\vspace{0.3cm}\includegraphics[width=0.5\textwidth]{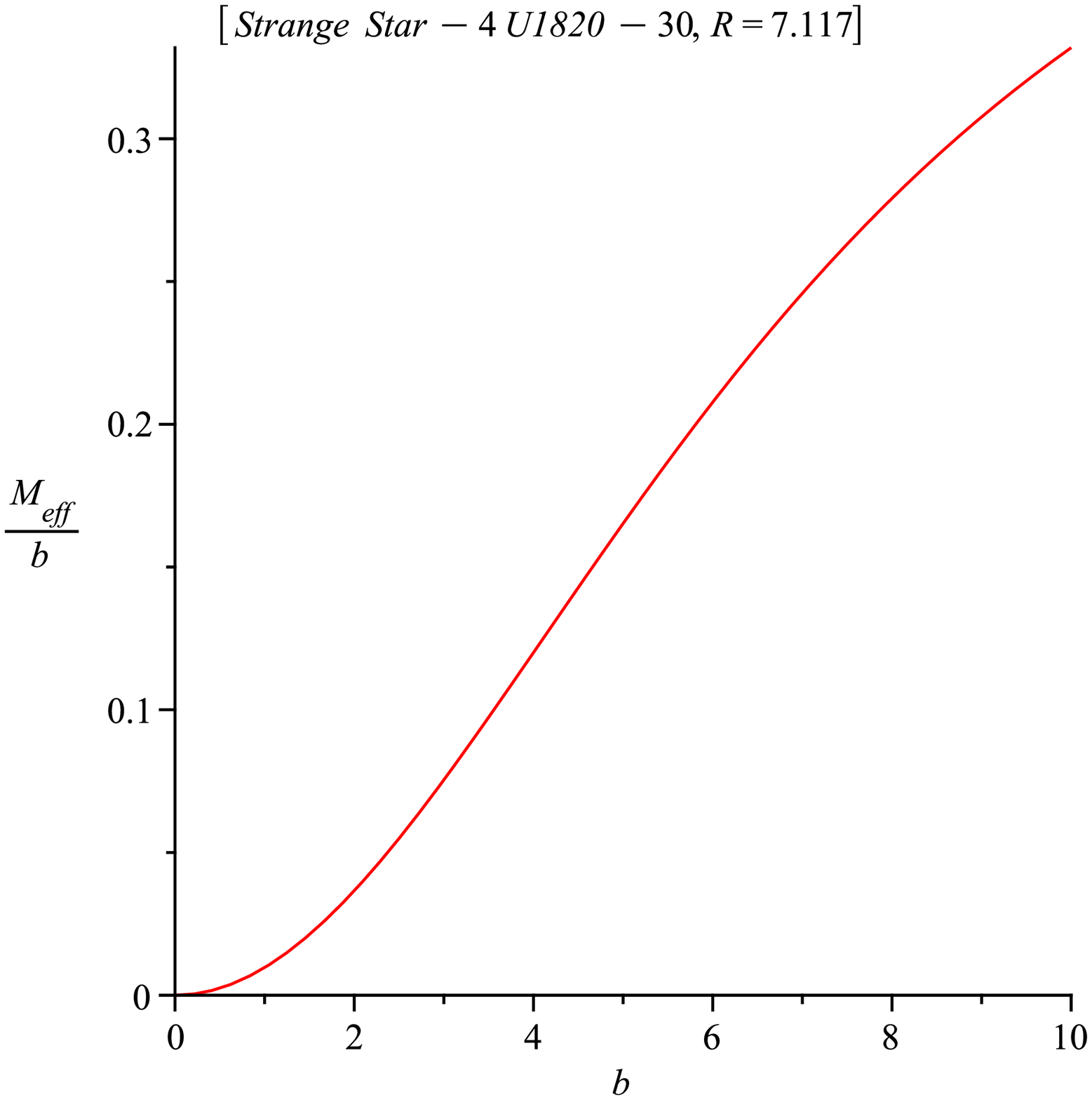}
\end{center}
\caption{Variation of $\frac{M_{eff}}{b}$ against radial parameter $r$.}
\label{fig:6}
\end{figure}

For a static spherically symmetric perfect fluid star in equilibrium, Buchdahl\cite{Buchdahl1959} showed that the maximum allowed mass-radius ratio is given by $\frac{2M}{R} < \frac{8}{9}$ (for a more generalized expression see Mak {\em et al}\cite{Mak2001}).  In our model, the effective gravitational mass is obtained  as
\begin{equation}
\label{eq31}
M_{eff} = 4\pi\int^{b}_{0} \rho r^2 dr =
 \frac{b}{2}\left[\frac{\frac{b^2}{R^2}}{1+\frac{b^2}{R^2}}\right].
\end{equation}
In Fig.~\ref{fig:5}, the effective mass for a given radius has been shown. The compactness of the star, accordingly, is given by
\begin{equation}
\label{eq32}
u= \frac{ M_{eff}(b)} {b}=
\frac{1}{2}\left[\frac{\frac{b^2}{R^2}}{1+\frac{b^2}{R^2}}\right],
\end{equation}
whose values at different radii have been shown in Fig.~\ref{fig:6}. We note that the constraint on the maximum allowed
mass-radius ratio in our case is similar to the isotropic fluid sphere, i.e., $\frac{M}{b} < \frac{4}{9}$. The corresponding surface redshift ($Z_s$) is obtained as
\begin{equation}
\label{eq33}
Z_s= \left[ 1-(2 u )\right]^{-\frac{1}{2}} - 1 = \sqrt{1+\frac{b^2}{R^2}}-1.
\end{equation}
The maximum surface redshift for different strange star candidates in this model has been shown in Table~$1$.

\begin{table}
\tbl{Values of the maximum surface redshift for different
strange star candidates.}
{\begin{tabular}{@{}cccc@{}} \toprule
Strange star candidate & $b$ (km) & $R$ (km) & $Z_s$(max) \\ \colrule
Her X-1 &  7.7 &  10.79 & 0.2285 \\
SAX J 1808.4-3658(SS1) &   7.07 &  5.787 & 0.5787 \\
SAX J 1808.4-3658(SS2) &  6.35 &   5.0282 & 0.6108 \\
4U 1820-30 & 10.0 &  7.117 & 0.7246 \\ \botrule
\end{tabular}}
\end{table}

\subsection{Estimated bag values of some strange star candidates}
Based on the analytic model developed so far, to get an estimate of the range of various physical parameters, let us now consider some strange star candidates like 4U 1820-30, Her X-1  and SAX J 1808.4-3658. Assuming the estimated mass and radius of these stars, we have calculated the values of the relevant physical parameters which have been compiled in Table $2$. In Table $3$, we have expressed the values of the bag constant, in particular, for respective stellar configurations.  We observe that a wide range of values of the bag constant are possible unless we impose some (yet unknown) constraints on $B$. However, from the perspective of a mathematically self-consistent model, it appears that a wide range of values of the bag constant are possible which is consistent with the CERN-SPS and RHIC data.

\begin{table}
\tbl{Values of the model parameters for different Strange
stars. (Data obtained for the star 4U 1820-30  have been
utilized to plot figures.) }
{\begin{tabular}{@{}ccccccccc@{}} \toprule
Strange star candidate & $M$ ($M_{\odot}$) & $b$ (km) &
$\frac{M}{b}$ & $R$ (km) & $B$ (km$^{-2}$)& $\rho_0$ (km$^{-2}$)
& $\rho_b$ (km$^{-2}$) & $p_0$ (km$^{-2}$) \\ \colrule
Her X-1 & 0.88 & 7.7 & 0.168 & 10.79 & 0.00013 & 0.00102 & 0.00052 & 0.000165\\
SAX J 1808.4-3658(SS1) & 1.435 & 7.07 & 0.299 & 5.787 & 0.00021 & 0.0035 & 0.00085 & 0.000901\\
SAX J 1808.4-3658(SS2) & 1.323 & 6.35 & 0.308 & 5.0282 & 0.00026 & 0.00472 & 0.00107 & 0.00121\\
4U 1820-30 & 2.25 & 10.0 & 0.332 & 7.117 & 0.0001 & 0.0023 & 0.00044 & 0.00063\\
\botrule
\end{tabular}}
\end{table}

\begin{table}
\tbl{Physical values of energy density, pressure and Bag
constant for different Strange stars.}
{\begin{tabular}{@{}ccccc@{}} \toprule
Strange star candidate &  Central density & Surface density & Central pressure & Bag constant\\
 & (gm~cm$^{-3}$) & (gm~cm$^{-3}$) & (dyne~cm$^{-2}$) & (MeV~fm$^{-3}$)\\ \colrule
 Her X-1 & $1.38169\times 10^{15}$ & $0.710012\times 10^{15}$ & $4.145070271\times 10^{35}$ & 99.7207866 \\
SAX J 1808.4-3658(SS1) & $4.808527413\times 10^{15}$ & $1.159209\times 10^{15}$ & $14.42558225\times 10^{35}$ & 162.8102741 \\
SAX J 1808.4-3658(SS2) & $6.370476078\times 10^{15}$ & $1.449092801\times 10^{15}$ & $19.11142825\times 10^{35}$ & 203.5242699 \\
4U 1820-30 & $3.179364432\times 10^{15}$ & $0.5960016695\times 10^{15}$ & $9.538093302\times 10^{35}$ & 83.70809967
\\
\botrule
\end{tabular}}
\end{table}

\section{Conclusion}
We have obtained a new class of solutions for the interior of a compact stellar object like a strange star. To construct the model, we have used the phenomenological MIT bag model EOS for quark matter. The analytic solution obtained is non-singular and is anisotropic in nature. The estimated values of the bag constant $B$ for different strange star candidates have been found to be on the higher side as compared to its estimated range of $60-80$~MeV~fm$^{-3}$ for a $\beta$-equilibrium stable strange matter configuration\cite{Farhi1984,Alcock1986}. It is likely that a wide range of values of the bag constant are permissible which is in agreement with the recent CERN-SPS and RHIC data.

\section*{Acknowledgments}
MK, AAU, RS  and FR  gratefully acknowledge support from IUCAA,
Pune, India, where a part of this work was carried out under its visiting research associateship programme. FR also gratefully acknowledges financial support from the PURSE, DST and UGC, Govt. of India.

\end{document}